Regular Article - Theoretical Physics

# Lyapunov exponent, ISCO and Kolmogorov–Senai entropy for Kerr–Kiselev black hole

**Monimala Mondal**[1,a], **Farook Rahaman**[1,b], **Ksh. Newton Singh**[1,2,c]

[1] Department of Mathematics, Jadavpur University, Kolkata, West Bengal 700032, India
[2] Department of Physics, National Defence Academy, Khadakwasla, Pune 411023, India

**Abstract** Geodesic motion has significant characteristics of space-time. We calculate the principle Lyapunov exponent (LE), which is the inverse of the instability timescale associated with this geodesics and Kolmogorov–Senai (KS) entropy for our rotating Kerr–Kiselev (KK) black hole. We have investigate the existence of stable/unstable equatorial circular orbits via LE and KS entropy for time-like and null circular geodesics. We have shown that both LE and KS entropy can be written in terms of the radial equation of innermost stable circular orbit (ISCO) for time-like circular orbit. Also, we computed the equation marginally bound circular orbit, which gives the radius (smallest real root) of marginally bound circular orbit (MBCO). We found that the null circular geodesics has larger angular frequency than time-like circular geodesics ($Q_o > Q_\sigma$). Thus, null-circular geodesics provides the fastest way to circulate KK black holes. Further, it is also to be noted that null circular geodesics has shortest orbital period ($T_{photon} < T_{ISCO}$) among the all possible circular geodesics. Even null circular geodesics traverses fastest than any stable time-like circular geodesics other than the ISCO.

## 1 Introduction

The first detection of black hole (BH) merger GW150914 strongly supports the existence of black holes [1,2] by observing the gravitational waves generated during the process of coalescence. Further, the first observation of the shadow of a super massive black hole in the giant elliptical galaxy M87 strengthen this claim for the existence of BHs [3]. BHs are usually surrounded by diffused matter in orbital motion named as "accretion disk". An accretion disk can be influenced by cosmic repulsion and magnetic field. A detail analysis was presented by Stuchlk et al. [4] by considering thin and thick disks. Stuchlk [5] have also discussed the motion of test particle with non-zero cosmological constant, further he also presented equation of motions for electrically charge and magnetic monopoles. There exist many stable and unstable circular orbits. Another type of orbit also exist which lie between dynamically unstable and stable orbits known as "homoclinic orbits" [6–8]. These orbits started asymptotically closed to unstable circular orbits and ends with spiraling in and out about the central BH. During the in-spiraling phase a significant amount of angular momentum is lost into gravitational waves thereby circularizing the orbit. In addition to these orbits, "chaotic orbits" have been discovered recently for fast spinning BHs [9–11]. The intermost stable circular orbits (ISCO) [12] is of more interest as it will identify the onset of dynamical instability.

The instability of an orbit can be identified by a positive value of Lyapunov exponent [13,14]. It is also well-known that maximal Lyapunov exponents (MLE) can distinguish a chaotic dynamics from others, which may initiate unstable/homoclinic orbits by a perturbation. For a spinning black hole the number of unstable orbits increase very rapidly and crowded into the corresponding phase [15–17]. However, determining LE in general relativistic regime is very difficult as each orbits has different LE and it is impractical to scan the overall nature of the orbits. Also, since LE determines the separation to neighboring orbits in time and time itself is relative which will eventually lead to relative LE [18]. Lyapunov coefficient delineates the measure of instability of the circular orbits and not a measure of chaos. On the other hand, if we consider magnetized black holes i.e. a black holes immersed in an external magnetic field, the equations of motion become chaotic [19–21]. Pánis et al. [22] discussed about Keplerian disk orbiting a Schwarzschild black hole embedded in an asymptotically uniform magnetic field where they have found three possible scenarios i.e. (1) regular oscillatory motion, (2) destruction due to capture by the magnetized black hole and (3) chaotic motion. They have used a time series of the solution of equations of motion for different conditions to analyzed the transition from regular to chaotic motion. Extending this work, Kološ et al. [23] explained observational data for few microquasars using magnetized standard geodesic models of QPOs. Further, Tursunov et al. [24] presents the possible source of ultrahigh-energy cosmic rays from supermassive black holes (SMBH), where the charged particles are accelerated by magnetic Penrose process from the rotating SMBH.

[a] e-mail: monimala.mondal88@gmail.com
[b] e-mail: rahaman@iucaa.ernet.in (corresponding author)
[c] e-mail: ntnphy@gmail.com

Wu et al. [25] developed a method of determining fast Lyapunov indicator (FLI) using two-nearby-orbits method without projection operations and with proper time as the independent variable, which can quickly identify chaos. Since LE is not invariant under coordinates transformations in GR, Wu and Huang [26] proposed a new relativistic LE which is invariant in a curve manifold. In general relativity, chaos takes an important role in describing more realistic physical systems. The amplification of chaos restricted relativistic three-body systems [27,28]. It has also been reported that chaos also exists in two relativistic systems like two fixed black holes [29,30], Schwarzschild's black hole and a dipolar shell [31–33] etc. Chaotic behavior must be well understood in binary gravitational waves sources or otherwise its detection will be highly uncertain. Levin et. al. [7] shown that in the absence of radiative back reaction a chaotic motion is developed in rapidly rotating compact stars where the relativistic precession of apastron is supplemented by chaotic precession of the orbital plane. Burd and Tavakol [34] used Bianchi type-IX cosmology to describe chaos as guage invariant model by calculation spectrum of LEs. The invariant formulation of LE was developed by Motter [35] showing that chaos is represented by positive LE and coordinate invariant.

The organization of the article is as follows: In Sect. 2: we begin with a basic definition of Lyapunov exponent and provide a simple formula for finding Lyapunov exponent ($\lambda$) in terms of the second derivative of the of effective potential in radial motion $\dot{r}^2$. In Sect. 3: We encapsulate the relation between Lyapunov exponent and KS-entropy. In Sect. 4: we analyze the rotating black hole and described completely the equatorial circular geodesics. We showed that Lyapunov exponent can be demonstrated in terms of ISCO equation and deliberated stability of time-like geodesics. We derived angular velocity and reciprocal of Critical exponent for both the cases of time-like and null-circular geodesics of the space-time. In Sect. 5: we constructed the equation of marginally bound circular orbit for finding the radius of marginally bound circular orbit to the black holes. We calculated the ratio of angular velocity of null-circular geodesics to time-like geodesics in Sect. 6. In Sect. 7: we determined the ratio of time period of null-circular geodesics to time-like geodesics. In Sect. 8: we summarize our work.

## 2 Proper time Lyapunov exponents and Radial potential

In a dynamical system the LE is a quantity that classified the rate of separation of extremely close trajectories. This rate of separation depends on orientations of initial separation vector. Among all LE, a positive Maximal Lyapunov exponent (MLE) is taken to indicate that the system is chaotic. If the system is conservative then the sum of all Lyapunov exponents must be zero and if negative then the system is dissipative. The Lyapunov spectrum can be utilized to give an evaluate of the rate of entropy production. According to Pesin's theorem [36], the sum of all positive LEs provide an estimate of the KS entropy. Also, a positive LE and a negative LE designates a divergence and a convergence between to nearby trajectories, respectively.

In classical physics, the study of autonomous smooth dynamical system analyzes differential equations of the form

$$\frac{dx}{dt} = F(x), \quad (1)$$

where $t$ represents the time parameter. If the following four conditions satisfied by chaos, we may quantify chaos in terms of LEs. The condition are as follows:

(i) the system is autonomous;
(ii) the relevant part of the phase space is bounded;
(iii) the invariant measure is normalizable;
(iv) the domain of the time parameter is infinite.

This types of characterization is advantageous for space-time because LE does not change under space diffeomorphisms of the form $z = \psi(x)$. That's why we may say that chaos is a property of the physical system and independent on the coordinate which is used to describe the system.

There is no absolute time in general relativity. Therefore, the time parameter in-force us to consider the Eq. (1) under the diffeomorphism: $z = \psi(x)$, $d\tau = \eta(x,t) dt$ in the space-time. Since the classical indicator of chaos i.e. LE and KS entropy depends on the choice of the time parameter, there is a conceptual problem. This problem was first arise in the mix-master cosmological model [37–39], where the largest LE was positive or zero depends on different choice of the coordinate. In general relativity (GR), this non-invariance features implies that chaos is not a property of physical system, it is a property of the coordinate system.

Following the work by Motter [35], we find that the chaos can be characterized by positive LE and KS entropy. The Lyapunov exponent and KS entropy transform under the space-time according to

$$\lambda_j^\tau = \frac{\lambda_j^t}{\langle \xi \rangle_t} \quad (j = 1, 2 \ldots, N), \quad (2)$$

and

$$h_{ks}^\tau = \frac{h_{ks}^t}{\langle \xi \rangle_t} \quad (3)$$

respectively. Where $N$ is phase-space dimension and $0 < \langle \xi \rangle_t < \infty$ is time average of $\xi = d\tau/dt$ over typical trajectory. For natural measure of the Eqs. (2) and (3) to be well defined, $0 < \langle \xi \rangle_t < \infty$ is the basic requirement condition. Finally we can say that, the coordinate transformation is always transformed into the time independent transformation which is given by

$$z = \psi(x), \quad d\tau = \xi(x,t)dt \quad (4)$$

where $\xi$ is a strictly positive, continuously differentiable function and $\psi$ is a diffeomorphism. Then it is clear that the LE and KS entropy are invariant under space diffeomorphism.

Now we will calculate the LE using proper time from

$$\lambda = \pm \sqrt{\frac{(\dot{r}^2)''}{2}}. \quad (5)$$

where $\dot{r}^2$ represents the radial potential. Here we will ignore the $\pm$ sign and we shall take only positive Lyapunov exponent. The circular orbit is stable when $\lambda$ is imaginary and circular orbit is unstable when $\lambda$ is real and when $\lambda = 0$, the circular orbit is marginally stable or saddle point.

Following Pretorius and Khurana [40], we can define critical exponent

$$\gamma = \frac{\Omega}{2\pi\lambda} = \frac{T_\lambda}{T_\Omega} \quad (6)$$

where, $T_\lambda$ represents the Lyapunov time scale, $T_\Omega$ represents the orbital time scale and $\Omega$ represent the angular velocity. Also here $T_\lambda = \frac{1}{\lambda}$ and $T_\Omega = \frac{2\pi}{\Omega}$. Now the critical exponent can be written, in terms of second order derivative of the square of radial velocity $(\dot{r}^2)$, as

$$\gamma = \frac{1}{2\pi}\sqrt{\frac{2\Omega^2}{(\dot{r}^2)''}}. \quad (7)$$

Also the reciprocal of critical exponent is given by

$$\frac{1}{\gamma} = \frac{T_\Omega}{T_\lambda} = 2\pi\sqrt{\frac{(\dot{r}^2)''}{2\Omega^2}}. \quad (8)$$

## 3 Lyapunov exponent and Kolmogorov–Senai entropy

Kolmogorov–Senai ($h_{ks}$) entropy [41,42] is an important quantity which is connected to the Lyapunov exponents. When the chaotic orbit evolves, this entropy gives a measurement about information lost or gained. Another way it can be explored that when $h_{ks} > 0$ then the system is chaotic or disorder and when $h_{ks} = 0$ then the system non-chaotic.

Following Pesin [43] the Kolmogorov–Senai ($h_{ks}$) entropy is the sum of all positive Lyapunov exponent, that is,

$$h_{ks} = \sum_{\lambda_j > 0} \lambda_j. \quad (9)$$

There are two Lyapunov exponent in 2-dimensional phase-space, so in terms of effective radial potential the Kolmogorov-senai entropy can be written as

$$h_{ks} = \sqrt{\frac{(\dot{r}^2)''}{2}} \quad (10)$$

When this entropy evolves with time it play an important role in dynamical system to examine whether a trajectory is in disorder or not. It is different from the statistical entropy or physical entropy. In fact, it assumes a partition of phase space.

## 4 Rotating Black hole

There are many choices of rotation axis and a multitude of angular momentum parameter in higher dimension whereas, only possible rotation axis with only one angular momentum for an axisymmetric in four dimension. Here we consider the metric of a Rotating black hole surrounding with quintessence [44] is,

$$ds^2 = -\left[1 - \frac{(2Mr + cr^{1-3w})}{\Sigma}\right]dt^2$$
$$+ \frac{\Sigma\,dr^2}{\Delta} - 2a\sin^2\theta\left[\frac{2Mr + cr^{1-3w}}{\Sigma}\right]dt\,d\phi + \Sigma\,d\theta^2$$
$$+ \sin^2\theta\left[r^2 + a^2 + a^2\sin^2\theta\left(\frac{2Mr + cr^{1-3w}}{\Sigma}\right)\right]d\phi^2. \quad (11)$$

where $c$ is a new parameter, which describes the hair of the black hole and $w$ represents a quintessential equation of state parameter and

$$\Delta = r^2 + a^2 - 2Mr - c \equiv (r - r_+)(r - r_-),$$
$$\Sigma = r^2 + a^2\cos^2\theta. \quad (12)$$

Also this rotating black hole is always bounded from above by $w = \frac{1}{3}$, which was first considered by Schee and Stuchlik [45]. Slaný a and Stuchlík [46] have also considered test particle in equatorial circular orbits around Kerr–Newman–de Sitter BH and naked singularity.

The metric is similar to Kerr–Newman black hole when $w = \frac{1}{3}, c = -Q^2$, Reissener-Nordström black hole when $w = \frac{1}{3}, c = -Q^2, a = 0$, Kerr black hole $w = \frac{1}{3}, c = 0$ and Schwarzschild black hole when $w = \frac{1}{3}, c = 0, a = 0$. The horizon take place at $g_{rr} = \infty$ or $\Delta = 0$ i.e

$$r_{\pm} = M \pm \sqrt{M^2 - a^2 + c}, \quad (13)$$

here $r_+$ is called event horizon and $r_-$ is called Cauchy horizon.

The BH solution considered in this work is the rotational version of the Kiselev static quintessence BH. Originally, Kiselev used quintessence surrounding the BH characterized by the EMT of the quintessence. Here, the scalar field couples to gravity through a Lagrangian of the form

$$\mathcal{L} = -\frac{1}{2} g^{\mu\nu} \partial_\mu \phi \, \partial_\nu \phi - V(\phi) \quad (14)$$

and the stress-energy tensor $T_{\mu\nu}$ for the matter field took as perfect fluid. The complete action and the field equations can be given as

$$S = \int d^2x \sqrt{-g} \left[ R - \frac{1}{2} g^{\mu\nu} \partial_\mu \phi \, \partial_\nu \phi - V(\phi) \right], \quad (15)$$

$$R_{\mu\nu} - \frac{1}{2} g_{\mu\nu} R = -8\pi T_{\mu\nu}$$
$$-8\pi \left[ \partial_\mu \phi \, \partial_\nu \phi - g_{\mu\nu} \left\{ \frac{1}{2} g^{\mu\nu} \partial_\mu \phi \, \partial_\nu \phi + V(\phi) \right\} \right], \quad (16)$$

$$\frac{1}{\sqrt{-g}} \partial_\mu \left[ \sqrt{-g} \, g^{\mu\nu} \partial_\nu \phi \right] - \frac{\partial V(\phi)}{\partial \phi} = 0. \quad (17)$$

The pressure and energy density of the scalar field are given by

$$p_\phi = \frac{1}{2} \dot\phi^2 - V(\phi) \;,\; \rho_\phi = \frac{1}{2} \dot\phi^2 + V(\phi). \quad (18)$$

The equation of state parameter $\omega_\phi = p_\phi/\rho_\phi$ is found to be

$$\omega_\phi = \frac{\dot\phi^2 - 2V(\phi)}{\dot\phi^2 + 2V(\phi)} \quad (19)$$

which lies between $-1 < \omega_\phi < 1$ and if $\dot\phi^2 < V(\phi)$ then $\omega_\phi < -1/3$. The rotating Kiselev BH was obtained by using Newman–Janis algorithm from Kiselev static quintessence BH, for details see [44].

### 4.1 Circular geodesics in the equatorial plane

To calculate the geodesic equation in the equatorial plane for this space-time we follow Chandrasekhar [47]. To compute the geodesics motions of the orbit in the equatorial plane we set $\dot\theta = 0$ and $\theta = $ constant $= \pi/2$. The appropriate Lagrangian for this motion is

$$2\mathcal{L} = -\left(1 - \frac{2M}{r} - cr^{-(1+3w)}\right)\dot t^2$$
$$-\left[\frac{4Ma}{r} + 2acr^{-(1+3w)}\right]\dot t\dot\phi$$
$$+\left[r^2 + \left(1 + \frac{2M}{r} + cr^{-(1+3w)}\right)a^2\right]\dot\phi^2 + \frac{r^2}{\Delta}\dot r^2, \quad (20)$$

where $\phi$ is an angular coordinate. Then the generalized momenta can be derived from it as follows

$$p_t = -\left[1 - \frac{2M}{r} - cr^{-(1+3w)}\right]\dot t$$
$$-\left[\frac{2Ma}{r} + acr^{-(1+3w)}\right]\dot\phi = -E = const. \quad (21)$$

$$p_\phi = -\left[\frac{2Ma}{r} + acr^{-(1+3w)}\right]\dot t$$
$$+\left[r^2 + \left(1 + \frac{2M}{r} + cr^{-(1+3w)}\right)a^2\right]$$
$$\dot\phi = L = const. \quad (22)$$

$$p_r = \frac{r^2}{\Delta}\dot r. \quad (23)$$

Since Lagrangian is unhampered by both of $t$ and $\phi$, so $p_t$ and $p_\phi$ preserve quantities. Solving the Eqs. (21) and (22) for $\dot\phi$ and $\dot t$ we get

$$\dot\phi = \frac{1}{\Delta}\left[\left(1 - \frac{2M}{r} - cr^{-(1+3w)}\right)L\right.$$
$$\left.+\left(\frac{2Ma}{r} + acr^{-(1+3w)}\right)E\right], \quad (24)$$

$$\dot t = \frac{1}{\Delta}\left[\left(r^2 + \left(1 + \frac{2M}{r} + cr^{-(1+3w)}\right)a^2\right)E\right.$$
$$\left.-\left(\frac{2Ma}{r} + acr^{-(1+3w)}\right)L\right]. \quad (25)$$

The Hamiltonian in terms of the metric is given by

$$2\mathcal{H} = 2p_t\dot t + 2p_\phi\dot\phi + 2p_r\dot r - 2\mathcal{L}$$
$$= -\left(1 - \frac{2M}{r} - cr^{-(1+3w)}\right)\dot t^2$$
$$-\left(\frac{4Ma}{r} + 2acr^{-(1+3w)}\right)\dot t\dot\phi$$
$$+\left[r^2 + \left(1 + \frac{2M}{r} + cr^{-(1+3w)}\right)a^2\right]\dot\phi^2$$
$$+\frac{r^2}{\Delta}\dot r^2 \quad (26)$$

Since the Hamiltonian does not depends on '$t$', then we can write it as follows:

$$2\mathcal{H} = -\left[\left(1 - \frac{2M}{r} - cr^{-(1+3w)}\right)\dot t\right.$$
$$+\left(\frac{2Ma}{r} + acr^{-(1+3w)}\right)\dot\phi\right]\dot t + \frac{r^2}{\Delta}\dot r^2$$
$$+\left[-\left(\frac{2Ma}{r} + acr^{-(1+3w)}\right)\dot t\right.$$
$$\left.+\left\{r^2 + \left(1 + \frac{2M}{r} + cr^{-(1+3w)}\right)a^2\right\}\dot\phi\right] \quad (27)$$

which is equivalent to

$$-E\dot{t} + L\dot{\phi} + \frac{r^2}{\Delta}\dot{r}^2 = \xi = constant \tag{28}$$

Here $\xi = -1, 0, 1$ for time-like geodesic, null geodesic and space-like geodesic, respectively. Putting Eqs. (24)–(25) in Eq. (28) we get

$$r^2\dot{r}^2 = r^2E^2 + \left(\frac{2M}{r} + cr^{-(1+3w)}\right)(aE - L)^2$$
$$+ (a^2E^2 - L^2) + \xi\Delta \tag{29}$$

### 4.1.1 Circular null geodesic

For circular null geodesics ($\xi = 0$), the radial Eq. (29) gives

$$r^2\dot{r}^2 = r^2E^2 + \left(\frac{2M}{r} + cr^{-(1+3w)}\right)(aE - L)^2$$
$$+ (a^2E^2 - L^2), \tag{30}$$

where, $E$ is the energy per unit mass and $L$ is the angular momentum per unit mass of the particle express the trajectory.

At the points $r = r_o$, $E = E_o$ and $L = L_o$ the circular geodesics condition $\dot{r}^2 = (\dot{r}^2)' = 0$ gives the following radius finding equations,

$$r_o^2 E_o^2 + \left(\frac{2M}{r_o} + cr_o^{-(1+3w)}\right)(aE_o - L_o)^2$$
$$+ (a^2E_o^2 - L_o^2) = 0, \tag{31}$$

$$2r_oE_o^2 - \left(\frac{2M}{r_o^2} + c(1+3w)r_o^{-(2+3w)}\right)(aE_o - L_o)^2 = 0. \tag{32}$$

We can write the above Eqs. (31) and (32) in simplified form by introducing the impact parameter $D_o = \frac{L_o}{E_o}$ as

$$r_o^2 + \left(\frac{2M}{r_o} + cr_o^{-(1+3w)}\right)(a - D_o)^2 + (a^2 - D_o^2) = 0, \tag{33}$$

$$r_o - \left(\frac{2M}{r_o^2} + \frac{c}{2}(1+3w)r_o^{-(2+3w)}\right)(a - D_o)^2 = 0. \tag{34}$$

From the Eq. (34) we get

$$D_o = a \mp \frac{r_o^2}{\sqrt{Mr_o + \frac{c}{2}(1+3w)r_o^{(1-3w)}}}. \tag{35}$$

Notice that the Eq. (33) is valid if and only if $|D_o| > a$. In the case of counter rotating orbits, $|D_o - a| = -(D_o - a)$, which correlates to upper sign in the above Eq. (35) while in the case of co-rotating orbits, $|D_o - a| = (D_o - a)$ which correlates to the lower sign in the above Eq. (35).

Putting the Eqs. (35) into (33), we obtain an equation for the radius of null circular geodesics

$$r_o^2 - 3Mr_o \pm 2a\sqrt{Mr_o + \frac{c}{2}(1+3w)r_o^{(1-3w)}}$$
$$- \frac{c}{2}(1+3w)r_o^{(1-3w)} = 0. \tag{36}$$

When $c = 0$, we find the well known result [47]. By using Eqs. (33) and (35) we can find an another important relation for null circular orbits as follows

$$D_o^2 = a^2 + r_o^2 \left(\frac{3Mr_o + \frac{3c}{2}(1+w)r_o^{(1-3w)}}{Mr_o + \frac{c}{2}(1+3w)r_o^{(1-3w)}}\right). \tag{37}$$

To analyze the null circular geodesics we will derived an important quantity is the angular frequency ($\Omega_\sigma$) which is given by

$$\Omega_o = \frac{\left(1 - \frac{2M}{r} - cr_o^{-(1+3w)}D_o\right) + a\left(\frac{2M}{r} + cr_o^{-(1+3w)}\right)}{\left(r^2 + a^2 + \frac{2Ma^2}{r} + a^2cr_o^{-(1+3w)}D_o\right) - a\left(\frac{2M}{r} + cr_o^{-(1+3w)}D_o\right)}$$
$$= \frac{1}{D_o} \tag{38}$$

With the help of the Eqs. (35) and (33) we show that the angular frequency ($\Omega_o$) of the equatorial null circular geodesics is inverse of the impact parameter $D_o$, which generalize the four dimensional result [47]. Also it is a general property of any stationary space-time.

### 4.1.2 Circular time-like geodesics

For time-like circular geodesics ($\xi = -1$), the radial Eq. (29) gives

$$r^2\dot{r}^2 = r^2E^2 + \left(\frac{2M}{r} + cr^{-(1+3w)}\right)(aE - L)^2$$
$$+ (a^2E^2 - L^2) - \Delta. \tag{39}$$

Now we shall focus on radial equation of ISCO which governing the circular time like geodesics in terms of reciprocal radius $u = \frac{1}{r}$, can be written as

$$\mathcal{V} = u^{-4}\dot{u}^2 = E^2 + 2Mu^3(aE - L)^2 + cu^{3(1+w)}(aE - L)^2$$
$$+ (a^2E^2 - L^2)u^2 - u^2(a^2 - cu^{3w-1}) + 2Mu - 1, \tag{40}$$

where $\mathcal{V}$ is the effective potential.

The condition for the existence of the circular orbits are at $r = r_\sigma$ or $u = u_\sigma$ is given by

$$\mathcal{V} = 0, \tag{41}$$

and

$$\frac{d\mathcal{V}}{du} = 0. \tag{42}$$

$\mathcal{V} = 0$ gives,

$$E^2 + 2Mu_\sigma^3(aE - L)^2 + cu_\sigma^{3(1+w)}(aE - L)^2$$
$$+ (a^2 E^2 - L^2)u_\sigma^2 - u_\sigma^2(a^2 - cu_\sigma^{3w-1}) + 2Mu_\sigma - 1 = 0, \quad (43)$$

and $\frac{d\mathcal{V}}{du} = 0$ gives,

$$3Mu_\sigma^2(aE - L)^2 + \frac{3c(1+w)}{2} u_\sigma^{2+3w}(aE - L)^2$$
$$+ (a^2 E^2 - L^2)u_\sigma - u_\sigma(a^2 - cu_\sigma^{3w-1})$$
$$+ \frac{c(3w-1)}{2} u_\sigma^{3w} + M = 0. \quad (44)$$

For circular orbits, let $L_\sigma$ and $E_\sigma$ be the value of energy and angular momentum at the radius $r_\sigma = \frac{1}{u_\sigma}$, respectively. Now putting $z = L_\sigma - aE_\sigma$, in the Eqs. (43) and (44) we obtain the followings equations

$$cu_\sigma^{3(1+w)} z^2 + 2Mu_\sigma^3 z^2 - (z^2 + 2azE_\sigma)u_\sigma^2$$
$$- u_\sigma^2(a^2 - cu_\sigma^{3w-1}) + 2Mu_\sigma - 1 + E_\sigma^2 = 0, \quad (45)$$

and

$$\frac{3c(1+w)}{2} u_\sigma^{(2+3w)} z^2 + 3Mu_\sigma^2 z^2 - (z^2 + 2azE_\sigma)u_\sigma$$
$$- u_\sigma(a^2 - cu_\sigma^{3w-1}) + \frac{c(3w-1)}{2} u_\sigma^{3w} + M = 0. \quad (46)$$

With the help of the Eq. (46) we get an equation for $E_\sigma^2$ from Eq. (45) as

$$E_\sigma^2 = 1 - u_\sigma M + Mz^2 u_\sigma^3 + \frac{c(3w-1)}{2} u_\sigma^{3w+1}$$
$$+ \frac{c(3w+1)}{2} u_\sigma^{3(1+w)} z^2. \quad (47)$$

Again from the Eq. (46) we get

$$2azE_\sigma u_\sigma = z^2 \left( 3Mu_\sigma^2 + \frac{3c(1+w)}{2} u_\sigma^{2+3w} - u_\sigma \right)$$
$$- \left( (a^2 - cu_\sigma^{3w-1})u_\sigma - \frac{c(3w-1)}{2} u_\sigma^{3w} - M \right). \quad (48)$$

After eliminating $E_\sigma$ from the Eqs. (47) and (48), we can get an quadratic equation for $z^2$ as

$$\mathcal{P} z^4 + \mathcal{Q} z^2 + \mathcal{S} = 0. \quad (49)$$

where

$$\mathcal{P} = u_\sigma^2 \left[ \left( 3Mu_\sigma - 1 + \frac{3c(1+w)}{2} u_\sigma^{3w+1} \right)^2 \right.$$
$$\left. - 4a^2 \left( Mu_\sigma^3 + \frac{c(1+3w)}{2} u_\sigma^{3(1+w)} \right) \right]$$

$$\mathcal{Q} = -2u_\sigma \left[ \left( 3Mu_\sigma - 1 + \frac{3c(1+w)}{2} u_\sigma^{3w+1} \right) \right.$$
$$\times \left( \left( (a^2 - cu_\sigma^{3w-1})u_\sigma - \frac{c(3w-1)}{2} u_\sigma^{3w} - M \right) \right.$$

$$+ 2a^2 u_\sigma \left( 1 - Mu_\sigma + \frac{c(3w-1)}{2} u_\sigma^{3w+1} \right) \right]$$

$$\mathcal{S} = \left[ \left( (a^2 - cu_\sigma^{3w-1})u_\sigma - \frac{c(3w-1)}{2} u_\sigma^{3w} - M \right) \right]^2$$

Now, the solution of the quadratic Eq. (49) is

$$z^2 = \frac{-\mathcal{Q} \pm \mathcal{D}}{2\mathcal{P}}, \quad (50)$$

where $D$ denotes the discriminant of the Eq. (49) which can be found as

$$D = 4au_\sigma \Delta_{u_\sigma} \sqrt{Mu_\sigma + \frac{c(3w+1)}{2} u_\sigma^{3w+1}}, \quad (51)$$

and

$$\Delta_{u_\sigma} = (a^2 - cu_\sigma^{3w-1})u_\sigma^2 - 2Mu_\sigma + 1. \quad (52)$$

In order to write the solution of the Eq. (49) in simple form, we consider the following expression

$$X_+ X_- = \left( 3Mu_\sigma - 1 + \frac{3c(1+w)}{2} u_\sigma^{3w+1} \right)^2$$
$$- 4a^2 \left( Mu_\sigma^3 + \frac{c(1+3w)}{2} u_\sigma^{3(1+w)} \right), \quad (53)$$

where

$$X_\pm = \left( 1 - 3Mu_\sigma - \frac{3c(1+w)}{2} u_\sigma^{3w+1} \right.$$
$$\left. \pm 2a \sqrt{Mu_\sigma^3 + \frac{c(3w+1)}{2} u_\sigma^{3(w+1)}} \right). \quad (54)$$

Thus the solution reduces to

$$z^2 u_\sigma^2 = \frac{\Delta_{u_\sigma} - X_\mp}{X_\mp} \quad (55)$$

Again we can utilize the identity

$$\Delta_{u_\sigma} - X_\mp = u_\sigma \left[ a\sqrt{u_\sigma} \pm \sqrt{M + \frac{c(3w+1)}{2} u_\sigma^{3w}} \right]^2 \quad (56)$$

Hence the solution for z can be written in the following simple form as

$$z = -\frac{\left[ a\sqrt{u_\sigma} \pm \sqrt{M + \frac{c(3w+1)}{2} u_\sigma^{3w}} \right]}{\sqrt{u_\sigma x_\pm}}. \quad (57)$$

Here the lower sign in the foregoing equations indicates to the co-rotating orbit whereas the upper sign indicates to the counter-rotating orbit. Inserting expression of (57) in the Eq. (47) we obtain following expression for $E_\sigma$ as energy

$$E_\sigma = \frac{1}{\sqrt{z_\pm}} \left[ 1 - 2Mu_\sigma \mp a\sqrt{Mu_\sigma + \frac{c(3w+1)}{2} u_\sigma^{3w+1}} \right] \quad (58)$$

Again with the help of Eqs. (57), (58) and the relation $L_\sigma = aE_\sigma + z$, we obtain the angular momentum for the time circular geodesics as

$$L_\sigma = \mp \frac{1}{\sqrt{u_\sigma z_\pm}} \left[ \sqrt{M + \frac{c(3w+1)}{2} u_\sigma^{3w}} \right.$$
$$\times \left( 1 + a^2 u_\sigma^2 \pm 2au_\sigma \sqrt{Mu_\sigma + \frac{c(3w+1)}{2} u_\sigma^{3w+1}} \right)$$
$$\left. \mp ac\sqrt{u_\sigma^5} \right] \quad (59)$$

In order to find the minimum radius for a stable circular orbit, we will obtained the second derivative of $\mathcal{V}$ with respect to $r$ for the value of $E_\sigma$ and $L_\sigma$ specific to circular orbit. Here we use the Eqs. (41, 42) with further equation

$$\left. \frac{d^2V}{du^2} \right|_{u=u_\sigma} = 0, \quad (60)$$

and we found that

$$\frac{d^2V}{du^2} = \frac{1}{u_\sigma} \left[ 6Mz^2 u_\sigma^2 + 3c(w+1)(3w+1)u_\sigma^{3w+2} z^2 \right.$$
$$\left. + 2c(3w-1)u_\sigma^{3w} - 2M \right] \quad (61)$$

By the Eq. (55) we get

$$\left. \frac{d^2V}{du^2} \right|_{u=u_\sigma} = \frac{2}{u_\sigma X_\pm} \left[ \left( 3M + \frac{3c(w+1)(3w+1)}{2} u_\sigma^{3w} \right) \Delta_{u_\sigma} \right.$$
$$\left. - \left( \frac{3c(w+1)(3w+1)}{2} u_\sigma^{3w} + c(3w-1)u_\sigma^{3w} + 4M \right) X_\pm \right] \quad (62)$$

Therefore the equation of ISCO at the reciprocal radius

$$\frac{2}{u_\sigma X_\pm} \left[ \left( 3M + \frac{3c(w+1)(3w+1)}{2} u_\sigma^{3w} \right) \Delta_{u_\sigma} \right.$$
$$\left. - \left( \frac{3c(w+1)(3w+1)}{2} u_\sigma^{3w} + c(3w-1)u_\sigma^{3w} + 4M \right) X_\pm \right] = 0 \quad (63)$$

or this is similar to

$$\left( 3Mu_\sigma^2 + \frac{3c(w+1)(3w+1)}{2} u_\sigma^{3w+2} \right)\left( a^2 - cu_\sigma^{3w-1} \right)$$
$$+ \frac{3}{4} c^2 (9w^2 + 18w + 1)(w+1) u_\sigma^{6w+1}$$
$$+ \frac{3cM}{2} (3w^2 + 14w + 3) u_\sigma^{3w+1}$$
$$\pm 8a \left( M + \frac{c}{8}(9w^2 + 18w + 1) u_\sigma^{6w} \right)$$
$$\times \sqrt{Mu_\sigma^3 + \frac{c(3w+1)}{2} u_\sigma^{3(w+1)}} + 6M^2 u_\sigma - M = 0 \quad (64)$$

Returning to the variable $r_\sigma$, we get the equation of innermost stable circular orbit(ISCO) for non extremal BH is identified by

$$Mr_\sigma^{6w+1} - 6M^2 r_\sigma^{6w} - 3Ma^2 r_\sigma^{6w-1}$$
$$- \frac{3cM}{2}(3w^2 + 14w + 3) r_\sigma^{3w} \pm 8a \left( Mr_\sigma^{3w} + \frac{c}{8}(9w^2 \right.$$
$$\left. + 18w + 1) \right) \sqrt{Mr_\sigma^{6w-1} + \frac{c(3w+1)}{2} r_\sigma^{3w-1}}$$
$$- 3c(w+1)\left[ \frac{a^2(3w+1)}{2} r_\sigma^{3w-1} \right.$$
$$\left. + \frac{c(9w^2 + 12w - 1)}{4} \right] = 0 \quad (65)$$

Let the smallest root of the above equation be $r_\sigma = r_{ISCO}$ which will be innermost stable circular orbit(ISCO) of the black hole. Here $(-)$ sign determines for direct orbit and $(+)$ sign determines for retrograde orbit.

Special cases:

(i) When $w = \frac{1}{3}$ and $c = -Q^2$, we can get the equation of ISCO for non-extremal Kerr–Newman black hole which is given by

$$Mr_\sigma^3 - 6M^2 r_\sigma^2 - 3Ma^2 r_\sigma + 9MQ^2 r_\sigma$$
$$\mp 8a(Mr_\sigma - Q^2)^{\frac{3}{2}} + 4Q^2(a^2 - Q^2) = 0. \quad (66)$$

The smallest root ( real ) of the Eq. (66) is the radius of ISCO.

(ii) When $w = \frac{1}{3}$ and $c = 0$, the Eq. (65) is similar to the equation of ISCO for Kerr black hole which is as follows:

$$r_\sigma^2 - 6Mr_\sigma \mp 8a\sqrt{Mr_\sigma} - 3a^2 = 0. \quad (67)$$

The radius of the ISCO is equal to the real root of the above equation.

(iii) When $w = \frac{1}{3}$ and $a = 0$, we obtained the equation of ISCO for Reissner-Nordström black hole as follows

$$Mr_\sigma^3 - 6M^2 r_\sigma^2 + 9MQ^2 r_\sigma - 4Q^4 = 0. \quad (68)$$

The radius of the ISCO for the Reissner-Nordström black hole can be found by obtaining the smallest real root of the previous equation.

(iv) When $w = \frac{1}{3}$ and $c = a = 0$, we obtain the radius of ISCO for Schwarzschild black hole is given by

$$r_\sigma - 6M = 0. \quad (69)$$

## 4.2 Lyapunov exponent

### 4.2.1 Time-like circular geodesics (equation of ISCO)

Now we will calculate the Lyapunov exponent and KS entry for time-circular orbit. Using the Eq. (5), we get the Lyapunov exponent and KS entry in terms of the radial equation of ISCO as follows

$$\lambda_{time} = h_{ks} = \sqrt{\frac{-\left(Mr_\sigma^{6w+1} - 6M^2 r_\sigma^{6w} - 3Ma^2 r_\sigma^{6w-1} - cMGr_\sigma^{3w} \pm 8aH - cT\right)}{r_\sigma^4 \left(r_\sigma^{3w+1} - 3Mr_\sigma^{3w} - \frac{3c(1+w)}{2} \pm 2a\sqrt{Mr_\sigma^{3w} + \frac{c(3w+1)}{2}}\right)}}, \tag{70}$$

where

$$G = \frac{3}{2}(3w^2 + 14w + 3),$$
$$H = \left(Mr_\sigma^{3w} + \frac{c}{8}(9w^2 + 18w + 1)\right)$$
$$\times \sqrt{Mr_\sigma^{6w-1} + \frac{c(3w+1)}{2}r_\sigma^{3w-1}},$$

$$T = 3(w+1)\left[\frac{a^2(3w+1)}{2}r_\sigma^{3w-1} + \frac{c(9w^2 + 12w - 1)}{4}\right]. \tag{71}$$

The condition for existing the time-circular geodesics motion of the test particle are energy($E_\sigma$) and angular momentum($L_\sigma$) must be real and finite. For these we must have $r_\sigma^{3w+1} - 3Mr_\sigma^{3w} - \frac{3c(1+w)}{2} \pm 2a\sqrt{Mr_\sigma^{3w} + \frac{c(3w+1)}{2}} > 0$ and $r_\sigma^{3w} > -\frac{c(3w+1)}{2M}$

From the above Eq. (70) we can conclude that the time-circular geodesics of non extremal black hole is stable when

$$Mr_\sigma^{6w+1} - 6M^2 r_\sigma^{6w} - 3Ma^2 r_\sigma^{6w-1}$$
$$-cMGr_\sigma^{3w} \pm 8aH - cT > 0, \tag{72}$$

that is, $\lambda_{time}$ or $h_{ks}$ is imaginary, the time-circular geodesics is unstable when

$$Mr_\sigma^{6w+1} - 6M^2 r_\sigma^{6w} - 3Ma^2 r_\sigma^{6w-1}$$
$$-cMGr_\sigma^{3w} \pm 8aH - cT < 0, \tag{73}$$

such that $\lambda_{time}$ or $h_{ks}$ is real and circular geodesics is marginally stable when

$$Mr_\sigma^{6w+1} - 6M^2 r_\sigma^{6w} - 3Ma^2 r_\sigma^{6w-1}$$
$$-cMGr_\sigma^{3w} \pm 8aH - cT = 0. \tag{74}$$

that is, $\lambda_{time}$ or $h_{ks}$ is zero.

Special cases:

(i) When $w = \frac{1}{3}, c = -Q^2$, we can get the Lyapunov exponent and KS entry for Kerr–Newman black hole in terms of ISCO equation are given by

$$\lambda_{KN} = h_{ks} = \sqrt{\frac{-\left(Mr_\sigma^3 - 6M^2 r_\sigma^2 - 3Ma^2 r_\sigma + 9MQ^2 r_\sigma \mp 8a(Mr_\sigma - Q^2)^{\frac{3}{2}} + 4Q^2(a^2 - Q^2)\right)}{r_\sigma^4 \left(r_\sigma^2 - 3Mr_\sigma \mp 2a\sqrt{Mr_\sigma - Q^2} + 2Q^2\right)}} \tag{75}$$

(ii) For Kerr black hole $w = \frac{1}{3}, c = 0$, the Lyapunov exponent and KS entry for circular time-like geodesics are

$$\lambda_{Kerr} = h_{ks}$$

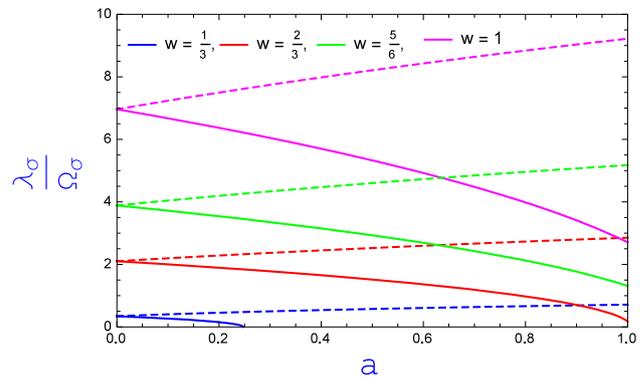

**Fig. 1** Dimensionless instability exponent $\frac{\lambda_\sigma}{\Omega_\sigma}$ as a function of rotation $a$ for real value of Lyapunov eponent. We use units such that $M = 1, c = 2$. Solid lines refer to corotating orbits and dashed lines refer to counterrotating orbits

$$= \sqrt{\frac{-M\left(r_\sigma^2 - 6M^2 r_\sigma \mp 8a\sqrt{Mr_\sigma} - 3a^2\right)}{r_\sigma^3\left(r_\sigma^2 - 3Mr_\sigma \mp 2a\sqrt{Mr_\sigma}\right)}} \quad (76)$$

(iii) For the Reissner-Nordström black hole $w = \frac{1}{3}, c = -Q^2, a = 0$, the Lyapunov exponent and KS entry for time-like geodesics are

$$\lambda_{RN} = h_{ks}$$
$$= \sqrt{\frac{-(Mr_\sigma^3 - 6M^2 r_\sigma^2 + 9MQ^2 r_\sigma - 4Q^4)}{r_\sigma^4\left(r_\sigma^2 - 3Mr_\sigma + 2Q^2\right)}} \quad (77)$$

(iv) For Schwarzschild black hole $w = \frac{1}{3}, c = -Q^2 = 0, a = 0$, the Lyapunov exponent and KS entry for circular time-like geodesics in terms of ISCO equation are given by

$$\lambda_{Sch} = h_{ks} = \sqrt{\frac{-M(r_\sigma - 6M)}{r_\sigma^3(r_\sigma - 3M)}} \quad (78)$$

*4.2.2 Null circular geodesics*

Lyapunov exponent and KS entry for circular null geodesics are

$$\lambda_{Null} = h_{ks}$$
$$= \sqrt{\frac{(L_o - aE_o)^2 \left(3Mr_o + 3c(1+w)r_o^{(1-3w)}\right)}{r_o^6}} \quad (79)$$

Since $(L_o - aE_o)^2 \geq 0$ and $r_o^{3w} > \frac{-c(w+1)}{M}$ then $\lambda_{Null}$ is real, so the circular null geodesics are unstable.

Special cases:

(i) When $w = \frac{1}{3}, c = -Q^2$, we can get the Lyapunov exponent and KS entry for Kerr–Newman black hole which is given by

$$\lambda_{KN} = h_{ks} = \sqrt{\frac{(L_o - aE_o)^2(3Mr_o - 4Q^2)}{r_o^6}}. \quad (80)$$

(ii) When $w = \frac{1}{3}, c = 0$, the Eq. (79) is similar to the equation Kerr black hole which is as follows :

$$\lambda_{Kerr} = h_{ks} = \sqrt{\frac{3M(L_o - aE_o)^2}{r_o^5}}. \quad (81)$$

(iii) For the Reissner-Nordström black hole $w = \frac{1}{3}, c = -Q^2, a = 0$, the Lyapunov exponent and KS entry for circular null geodesics are

$$\lambda_{RN} = h_{ks} = \sqrt{\frac{L_o^2(3Mr_o - 4Q^2)}{r_o^6}}. \quad (82)$$

When $r_o > \frac{4}{3}\frac{Q^2}{M}$, $\lambda_{RN}$ is real which implies that null-circular geodesics for Reissner- Nordström black hole is unstable.

(iv) For Schwarzschild black hole $w = \frac{1}{3}, c = 0, a = 0$, the Lyapunov exponent and KS entry for circular null geodesics are given by

$$\lambda_{Sch} = h_{ks} = \sqrt{\frac{3ML_o^2}{r_\sigma^5}}. \quad (83)$$

It can be verify that when $r_o = 3M$, $\lambda_{Sch}$ is real. Hence null-circular geodesics For Schwarzschild photon sphere are unstable.

4.3 Angular velocity of time-like circular geodesic

For time-like circular geodesic, the angular velocity at $r = r_\sigma$ is given by

$$\Omega_\sigma = \frac{\dot{\phi}}{\dot{t}} = \frac{\left(1 - \frac{2M}{r_\sigma} - cr_\sigma^{-(1+3w)}\right)L_\sigma + a\left(\frac{2M}{r_\sigma} + cr_\sigma^{-(1+3w)}\right)aE_\sigma}{\left(r_\sigma^2 + a^2 + \frac{2Ma^2}{r_\sigma} + a^2 cr_\sigma^{-(1+3w)}\right)E_\sigma - a\left(\frac{2M}{r_\sigma} + cr_\sigma^{-(1+3w)}\right)aL_\sigma} \quad (84)$$

The Eq. (84) can be written as

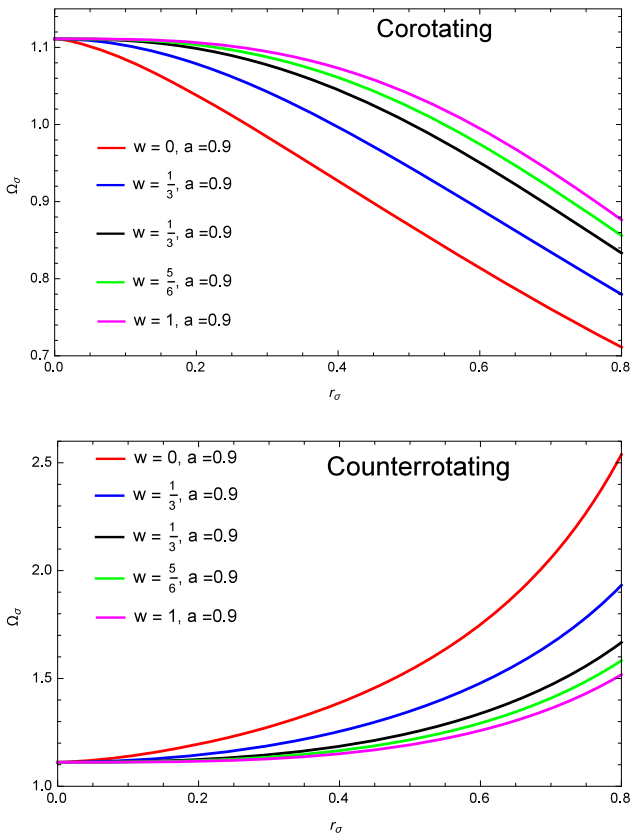

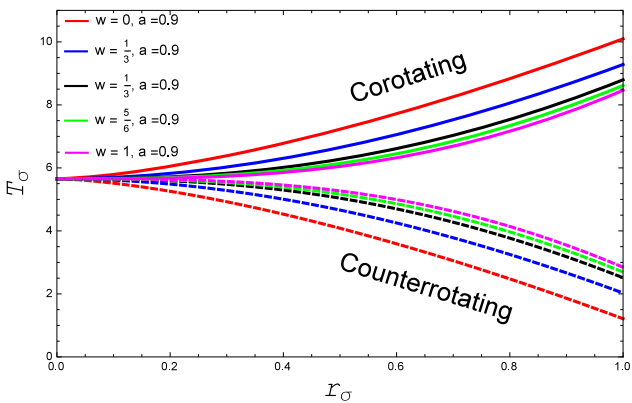

**Fig. 2** The plots shows the time-like orbital frequency $\Omega_\sigma$ to the radial coordinate $r_\sigma$ of corotating orbits (left panel) and counterrotating orbits (lower panel) for different values of $w$. We use units $M = 1$, $a = 0.9$ and $c = 2$

**Fig. 3** The plot shows the orbital time period of time-like circular geodesic $T_\sigma$ to the radius $r_\sigma$. We use units such that $M = 1$, $a = 0.9$ and $c = 2$. Solid lines refer to corotating orbits and dashed lines refer to counterrotating orbits

$$\Omega_\sigma = \frac{[L_\sigma - 2Mu_\sigma z - cu_\sigma^{(1+3w)}z]u_\sigma^2}{(1+a^2 u_\sigma^2)E_\sigma - 2Mau_\sigma^3 z - acu_\sigma^{3(1+3)}z}. \quad (85)$$

The numerator of the Eq. (85) can be written in the simplified form as

$$L_\sigma - 2Mu_\sigma z - cu_\sigma^{(1+3w)}z = \mp \frac{\sqrt{M + \frac{c(3w+1)}{2}u_\sigma^{3w}}}{\sqrt{u_\sigma z_\pm}}\Delta_{u_\sigma}. \quad (86)$$

Similarly, the denominator can be written as

$$(1+a^2 u_\sigma^2)E_\sigma - 2Mau_\sigma^3 z - acu_\sigma^{3(1+3)}z$$
$$= \frac{\Delta_{u_\sigma}}{\sqrt{z_\pm}}\left(1 \mp a\sqrt{Mu_\sigma^3 + \frac{c(3w+1)}{2}u_\sigma^{3(w+1)}}\right). \quad (87)$$

Putting Eqs. (86) and (87) into Eq. (85) we obtain the angular velocity for time circular orbit as

$$\Omega_\sigma = \mp \frac{\sqrt{Mu_\sigma^3 + \frac{c(3w+1)}{2}u_\sigma^{3(w+1)}}}{1 \mp a\sqrt{Mu_\sigma^3 + \frac{c(3w+1)}{2}u_\sigma^{3(w+1)}}}. \quad (88)$$

Now the angular velocity in terms of $r_\sigma$ for time circular geodesics is

$$\Omega_\sigma = \mp \frac{\sqrt{Mr_\sigma^{3w} + \frac{c(3w+1)}{2}}}{r_\sigma^{\frac{3(w+1)}{2}} \mp a\sqrt{Mr_\sigma^{3w} + \frac{c(3w+1)}{2}}}. \quad (89)$$

Now the time period for circular time-like orbit is given by

$$T_\sigma = \frac{2\pi}{\Omega_\sigma} = \mp 2\pi \cdot \frac{r_\sigma^{\frac{3(w+1)}{2}} \mp a\sqrt{Mr_\sigma^{3w} + \frac{c(3w+1)}{2}}}{\sqrt{Mr_\sigma^{3w} + \frac{c(3w+1)}{2}}} \quad (90)$$

At the limit $w = \frac{1}{3}$, $c = 0$, $a = 0$, the above equation can be written in the form $T_\sigma^2 \propto r_\sigma^3$. This shows that the Eq. (90) satisfies relativistic Kepler's law for Schwarzschild black hole.

### 4.4 Critical exponent

#### 4.4.1 Time-like circular geodesics

Now we will calculate the critical exponent for equatorial time-like circular geodesics. Thus the reciprocal of critical exponent is followed by the Eq. (8)

$$\frac{1}{\gamma} = 2\pi \sqrt{\frac{-\left(Mr_\sigma^{6w+1} - 6M^2 r_\sigma^{6w} - 3Ma^2 r_\sigma^{6w-1} - cMGr_\sigma^{3w} \pm 8aH - cT\right)\left(r_\sigma^{\frac{3(w+1)}{2}} \mp a\sqrt{Mr_\sigma^{3w} + \frac{c(3w+1)}{2}}\right)^2}{r_\sigma^4 \left(Mr_\sigma^{3w} + \frac{c(3w+1)}{2}\right)\left(r_\sigma^{3w+1} - 3Mr_\sigma^{3w} - \frac{3c(1+w)}{2} \pm 2a\sqrt{Mr_\sigma^{3w} + \frac{c(3w+1)}{2}}\right)}} \quad (91)$$

Since $\left(r_\sigma^{\frac{3(w+1)}{2}} \mp a\sqrt{Mr_\sigma^{3w} + \frac{c(3w+1)}{2}}\right)^2 \geq 0$, $\left(r_\sigma^{3w+1} - 3Mr_\sigma^{3w} - \frac{3c(1+w)}{2} \pm 2a\sqrt{Mr_\sigma^{3w} + \frac{c(3w+1)}{2}}\right) \geq 0$, $r_\sigma^{3w} > -\frac{c(3w+1)}{2M}$ and $Mr_\sigma^{6w+1} - 6M^2 r_\sigma^{6w} - 3Ma^2 r_\sigma^{6w-1} - cMGr_\sigma^{3w} \pm 8aH - cT < 0$, so $\frac{1}{\gamma}$ is real which shows that equatorial time-like circular geodesics is unstable.

Special cases:

(i) When $w = \frac{1}{3}, c = -Q^2$, we can get the equation of critical exponent for Kerr–Newman black hole in terms of ISCO which is given by

$$\frac{1}{\gamma} = 2\pi \sqrt{\frac{-(Mr_\sigma^3 - 6M^2 r_\sigma^2 - 3Ma^2 r_\sigma + 9MQ^2 r_\sigma \mp 8a(Mr_\sigma - Q^2)^{\frac{3}{2}} + 4Q^2(a^2 - Q^2)(r_\sigma^2 \mp a\sqrt{Mr_\sigma - Q^2})^2}{r_\sigma^4 (Mr_\sigma - Q^2)(r_\sigma^2 - 3Mr_\sigma \mp 2a\sqrt{Mr_\sigma - Q^2} + 2Q^2)}} \quad (92)$$

(ii) When $w = \frac{1}{3}$ and $c = 0$, the Eq. (91) is similar to the equation of Critical exponent for Kerr black hole in terms of ISCO which is as follows:

$$\frac{1}{\gamma} = 2\pi \sqrt{\frac{-\left(r_\sigma^2 - 6M^2 r_\sigma \mp 8a\sqrt{Mr_\sigma} - 3a^2\right)(r_\sigma\sqrt{r_\sigma} \mp a\sqrt{M})^2}{r_\sigma^3 \left(r_\sigma^2 - 3Mr_\sigma \mp 2a\sqrt{Mr_\sigma}\right)}} \quad (93)$$

(iii) For the Reissner-Nordström black hole $w = \frac{1}{3}, c = -Q^2$ and $a = 0$, the reciprocal of Critical exponent for time-like geodesic is

$$\frac{1}{\gamma} = 2\pi \sqrt{\frac{-(Mr_\sigma^3 - 6M^2 r_\sigma^2 + 9MQ^2 r_\sigma - 4Q^4)}{(Mr_\sigma - Q^2)(r_\sigma^2 - 3Mr_\sigma + 2Q^2)}} \quad (94)$$

(iv) When $w = \frac{1}{3}$ and $c = a = 0$, we obtained the reciprocal of Critical exponent for Schwarzschild black hole in terms of ISCO equation

$$\frac{1}{\gamma} = 2\pi \sqrt{\frac{-(r_\sigma - 6M)}{r_\sigma - 3M}} \quad (95)$$

*4.4.2 Null circular geodesics*

The reciprocal of critical exponent associated to null circular geodesics is given by

$$\left(\frac{1}{\gamma}\right)_{Null} = 2\pi \left(r_o^{\frac{3(w+1)}{2}} \mp a\sqrt{Mr_o^{3w} + \frac{c(3w+1)}{2}}\right)$$
$$\times \sqrt{\frac{(L_o - aE_o)^2 \left(3Mr_o + 3c(1+w)r_o^{(1-3w)}\right)}{r_o^6 \left(Mr_o^{3w} + \frac{c(3w+1)}{2}\right)}} \quad (96)$$

Special cases:
We take following limits in the above Eq. (96):

(i) When $w = \frac{1}{3}$ and $c = -Q^2$, we can get reciprocal of Critical exponent for Kerr–Newman black hole which is given by

$$\left(\frac{1}{\gamma}\right)_{Null} = 2\pi \sqrt{\frac{(L_o - aE_o)^2(3Mr_o - 4Q^2)(r_o^2 \mp a\sqrt{Mr_o - Q^2})^2}{r_o^6(Mr_o - Q^2)}} \quad (97)$$

(ii) When $w = \frac{1}{3}$ and $c = 0$, the Eq. (96) is similar to the reciprocal of Critical exponent of null-circular geodesics for Kerr black hole which is as follows:

$$\left(\frac{1}{\gamma}\right)_{null} = 2\pi \sqrt{\frac{3(L_o - aE_o)^2(r_o\sqrt{r_o} \mp a\sqrt{M})^2}{r_0^5}} \quad (98)$$

(iii) For the Reissner-Nordström black hole $w = \frac{1}{3}, c = -Q^2$ and $a = 0$, the reciprocal of Critical exponent for null-circular geodesic is

$$\left(\frac{1}{\gamma}\right)_{null} = 2\pi \sqrt{\frac{L_o^2(3Mr_o - 4Q^2)}{r_o^2(Mr_o - Q^2)}} \quad (99)$$

(iv) When $w = \frac{1}{3}$ and $c = a = 0$, we obtained the reciprocal of Critical exponent for Schwarzschild black hole in terms of null-circular geodesics is

$$\left(\frac{1}{\gamma}\right)_{null} = 2\pi \sqrt{\frac{3L_o^2}{r_o^2}} \quad (100)$$

## 5 Marginally bound circular orbit

It is known that for stable circular orbits the effective potential has a the local minima. Thus the condition for the existence of stable circular orbit exists is $d^2V_{eff}/dr^2 > 0$. However, for marginally stable circular orbits the condition $d^2V_{eff}/dr^2 = 0$ should be satisfied. For marginally bound circular orbit we have $E_\sigma = 1$ or $E_\sigma^2 = 1$ however, this is not true for asymptotically non-flat spacetimes [48].

Now from Eq. (47) we get

$$z^2 u_\sigma^2 = \frac{Mu_\sigma - \frac{c(3w+1)}{2}u_\sigma^{3w+1}}{Mu_\sigma + \frac{c(3w+1)}{2}u_\sigma^{3w+1}}, \quad (101)$$

Again from Eq. (57) we have

$$z^2 u_\sigma^2 = \frac{\left[au_\sigma \pm \sqrt{Mu_\sigma + \frac{c(3w+1)}{2}u_\sigma^{3w+1}}\right]^2}{x_\mp}. \quad (102)$$

Hence from (101) and (102) we get

$$x_\mp = \left[\frac{Mu_\sigma + \frac{c(3w+1)}{2}u_\sigma^{3w+1}}{Mu_\sigma - \frac{c(3w-1)}{2}u_\sigma^{3w+1}}\right]$$
$$\times \left[au_\sigma \pm \sqrt{Mu_\sigma + \frac{c(3w+1)}{2}u_\sigma^{3w+1}}\right]^2, \quad (103)$$

or the above equation can be written as

$$\left(Mu_\sigma - \frac{c(3w-1)}{2}u_\sigma^{3w+1}\right)\left(1 - 3Mu_\sigma - \frac{3c(1+w)}{2}u_\sigma^{3w+1}\right)$$
$$\pm 2a\sqrt{Mu_\sigma^3 + \frac{c(3w+1)}{2}u_\sigma^{3(w+1)}}\bigg)$$
$$= \left(Mu_\sigma + \frac{c(3w+1)}{2}u_\sigma^{3w+1}\right)$$
$$\times \left[au_\sigma \pm \sqrt{Mu_\sigma + \frac{c(3w+1)}{2}u_\sigma^{3w+1}}\right]^2 \quad (104)$$

After simplification we get the equation of marginally bound circular orbit in terms of $u_\sigma$ as follows:

$$\left(a^2 + \frac{c}{2}(3w+1)u_\sigma^{3w-1} + \frac{3(3w-1)(w+1)}{2(3w+1)}u_\sigma^{3w-1}\right)$$
$$-M\left(\frac{c}{2}(9w+5)u_\sigma^{3w-1} + a^2 - \frac{3c}{2}(3w-1)u_\sigma^{3w-1}\right)u_\sigma^2$$
$$-\frac{c}{2}(3w+1)u_\sigma^{3w+2} - \left(4M^2u_\sigma + \frac{c}{2}(3w-1)u_\sigma^{3w}\right)$$
$$\mp \left(4aMu_\sigma + ac(3w+1)u_\sigma^{3w-1}\right)$$
$$\times \sqrt{Mu_\sigma + \frac{c(3w+1)}{2}u_\sigma^{(3w+1)}} + M = 0 \quad (105)$$

Now reverting to $r_\sigma$ we can get the above equation as

$$Mr_\sigma^{6w+1} - \left(4M^2 r_\sigma^{6w} + \frac{c}{2}(3w-1)r_\sigma^{3w+1}\right)$$
$$-M\left(\frac{c}{2}(9w+5)r_\sigma^{3w} + a^2 r_\sigma^{6w-1} - \frac{3c}{2}(3w-1)r_\sigma^{3w}\right)$$
$$\mp \left(4aMr_\sigma^{3w} + ac(3w+1)\right)\sqrt{Mr_\sigma^{6w-1} + \frac{c(3w+1)}{2}r_\sigma^{(3w-1)}}$$
$$-\left(a^2 r_\sigma^{3w-1} + \frac{c}{2}(3w+1) + \frac{3(3w-1)(w+1)}{2(3w+1)}\right) = 0. \quad (106)$$

Let $r_\sigma = r_{mb}$ be the smallest(real) root of the Eq. (106). The root will be the closest bound circular orbit to the black hole.

Special cases:

(i) When $w = \frac{1}{3}$ and $c = -Q^2$, we can get the equation of marginally bound circular orbit for Kerr–Newman black hole which is given by

$$Mr_\sigma^3 - 4M^2 r_\sigma^2 - Ma^2 r_\sigma + 4MQ^2 r_\sigma \\ \mp (4aMr_\sigma - 2aQ^2)\sqrt{Mr_\sigma - Q^2} + Q^2(a^2 - Q^2) = 0. \quad (107)$$

The smallest root (real) of the Eq. (107) is the radius of marginally bound circular orbit of the black hole.

(ii) When $w = \frac{1}{3}$ and $c = 0$, the Eq. (106) is similar to the equation of Marginally bound circular orbit for Kerr black hole which is as follows :

$$r_\sigma^2 - 4Mr_\sigma \mp 4a\sqrt{Mr_\sigma} - a^2 = 0. \quad (108)$$

The radius say $r_\sigma = r_{mb}$ can be obtained by finding real smallest root of the above equation for the marginally bound circular orbit of the black hole.

(iii) For the Reissner-Nordström black hole $w = \frac{1}{3}, c = -Q^2$ and $a = 0$, we can get the equation of marginally bound circular orbit which is as follows

$$Mr_\sigma^3 - 4M^2 r_\sigma^2 + 4MQ^2 r_\sigma - Q^4 = 0 \quad (109)$$

The radius of marginally bound circular orbit of the black hole can be found by obtaining the smallest real root of the above equation.

(iv) When $w = \frac{1}{3}$ and $c = a = 0$, we can obtained the radius of marginally bound circular orbit for Schwarzschild back hole as follows

$$r_\sigma - 4M = 0. \quad (110)$$

More specifically, we can say that $r_\sigma = 4M$ is the radius of marginally bound circular orbit for Schwarzschild back hole.

The static radius for the circular orbits of test particles is given by the condition $\frac{d^2 V_{eff}}{dr^2} = 0$ i.e. correspond to local extrema of the effective potential [49]. Actually, the radii of circular orbits are restricted by the static radius. The important notion of the static radius will be discussed in details in a future project.

## 6 Ratio of angular velocity between null-circular geodesics and time-circular geodesics

We have already computed the angular velocity for time circular geodesics [50] in the Eq. (89) which is given by

$$\Omega_\sigma = \mp \frac{\sqrt{Mr_\sigma^{3w} + \frac{c(3w+1)}{2}}}{r_\sigma^{\frac{3(w+1)}{2}} \mp a\sqrt{Mr_\sigma^{3w} + \frac{c(3w+1)}{2}}}. \quad (111)$$

Again, the similar expression of angular velocity $\Omega_o = \frac{1}{D_o}$ for null circular geodesics can be obtain from the above equation as

$$\Omega_o = \mp \frac{\sqrt{Mr_o^{3w} + \frac{c(3w+1)}{2}}}{r_o^{\frac{3(w+1)}{2}} \mp a\sqrt{Mr_o^{3w} + \frac{c(3w+1)}{2}}}. \quad (112)$$

Now the ratio of angular velocity between null-circular geodesics and time-circular geodesics is

$$\frac{\Omega_o}{\Omega_\sigma} = \left(\frac{\sqrt{Mr_o^{3w} + \frac{c(3w+1)}{2}}}{\sqrt{Mr_\sigma^{3w} + \frac{c(3w+1)}{2}}}\right) \\ \times \left(\frac{r_\sigma^{\frac{3(w+1)}{2}} \mp a\sqrt{Mr_\sigma^{3w} + \frac{c(3w+1)}{2}}}{r_o^{\frac{3(w+1)}{2}} \mp a\sqrt{Mr_o^{3w} + \frac{c(3w+1)}{2}}}\right). \quad (113)$$

(i) When the radius of time-circular geodesics is equal to the radius of null circular geodesic, the corresponding angular velocities are also equal, that is, when $r_\sigma = r_o$, $Q_\sigma = Q_o$ which shows that the intriguing physical phenomena could occurs in the curve space-time. For instance, it would increase the interesting possibility of exciting quasinormal frequencies of the black hole by orbiting particle, possibly leading to instabilities of the space-time [51,52].

(ii) When the radius of time-circular geodesics is greater than the radius of null-circular geodesics, that is, $r_\sigma > r_o$ then $Q_o > Q_\sigma$ which implies that null-circular geodesics is characterized by the largest angular frequency than the angular frequency of time-circular geodesics as measured by asymptotic observer. Therefore, such type of space-time are characterized by

$$\Omega_{null} > \Omega_{timelike}. \quad (114)$$

This type of characteristic of null-circular geodesic has been graphed in the Fig. 4 for different values of radius $r$.

Thus from the above equation we may conclude that the null-circular geodesic provide the fastest way to circle of black hole [53,54]. This is satisfied for the case of Spherically symmetry Schwarzschild black hole, Kerr black

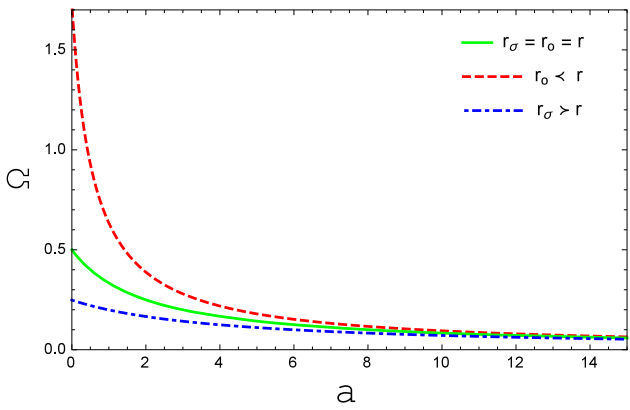

**Fig. 4** The plot shows the orbital angular frequency $\Omega$ Versus $a$ for different values of radius $r$ in the range $r_o \leq r \leq r_\sigma$, where $r_\sigma$ and $r_o$ are the radius of time-like geodesic and null-circular geodesic, respectively. Green (solid) color indicates for the value where the radius of time-like and null-circular geodesics are coincide, Red(Dashed) color indicates for the value of radius of null-circular geodesic and blue (Dot-Dashed) color indicates the value of radius of time-like geodesic. We use units such that $M = 1, c = 2$

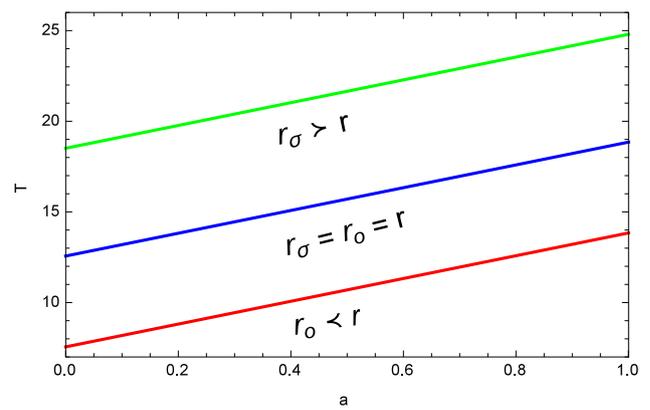

**Fig. 5** The plot shows the orbital time period $T$ versus $a$ for different values of radius $r$ in the range $r_o \leq r \leq r_\sigma$, where $r_\sigma$ and $r_o$ are the radius of time-like geodesic and null-circular geodesic, respectively. Blue color indicates for the value where the radius of time-like and null-circular geodesics are coincide, green color indicates for the value of radius of time-like geodesic and red color indicates the value of radius of null-circular geodesic. We use units such that $M = 1, c = 2$

hole and is still much general and can be applied for the case of Stationary, Kerr–Newman, axisymmetry spacetime.

## 7 Ratio of Time period between null-circular geodesics and time-circular geodesics

From the Eq. (90), the time period for time-like geodesics is given by

$$T_\sigma = \mp 2\pi \frac{\left(r_\sigma^{\frac{3(w+1)}{2}} \mp a\sqrt{Mr_\sigma^{3w} + \frac{c(3w+1)}{2}}\right)}{\sqrt{Mu_\sigma^{3w} + \frac{c(3w+1)}{2}}}. \quad (115)$$

Substituting $r_\sigma = r_o$ in the above equation, we can deduce the time period for null-circular geodesics as

$$T_o = \mp 2\pi \frac{\left(r_o^{\frac{3(w+1)}{2}} \mp a\sqrt{Mr_o^{3w} + \frac{c(3w+1)}{2}}\right)}{\sqrt{Mu_o^{3w} + \frac{c(3w+1)}{2}}}. \quad (116)$$

Now the ratio of time period between null-circular geodesics and time-circular geodesics is given by

$$\frac{T_o}{T_\sigma} = \left(\frac{\sqrt{Mr_\sigma^{3w} + \frac{c(3w+1)}{2}}}{\sqrt{Mr_o^{3w} + \frac{c(3w+1)}{2}}}\right)$$

$$\times \left(\frac{r_o^{\frac{3(w+1)}{2}} \mp a\sqrt{Mr_o^{3w} + \frac{c(3w+1)}{2}}}{r_\sigma^{\frac{3(w+1)}{2}} \mp a\sqrt{Mr_\sigma^{3w} + \frac{c(3w+1)}{2}}}\right). \quad (117)$$

(i) When $r_\sigma = r_o$, the Eq. (115) and the Eq. (116) became identical, that is, the time period for time-like geodesics is similar to the time period of null-circular geodesics which leading to excitations of quasinormal modes.
(ii) When the radius of null-circular geodesics is smaller than the radius of time-like geodesics, orbital time period of null-circular geodesics is smaller than the orbital time period of time-circular geodesics, that is, for $r_o < r_\sigma$, $T_o < T_\sigma$. Let $r_o = r_{photon}$, $r_\sigma = r_{ISCO}$, the ratio between the orbital time period for photon sphere and the orbital time period for ISCO be

$$\frac{T_{photon}}{T_{ISCO}} = \left(\frac{\sqrt{Mr_\sigma^{3w} + \frac{c(3w+1)}{2}}}{\sqrt{Mr_o^{3w} + \frac{c(3w+1)}{2}}}\right)$$

$$\times \left(\frac{r_o^{\frac{3(w+1)}{2}} \mp a\sqrt{Mr_o^{3w} + \frac{c(3w+1)}{2}}}{r_\sigma^{\frac{3(w+1)}{2}} \mp a\sqrt{Mr_\sigma^{3w} + \frac{c(3w+1)}{2}}}\right). \quad (118)$$

which leads to

$$T_{photon} < T_{ISCO}, \quad (119)$$

this implies that the time-like circular geodesics provides the slowest possible orbital time period. This type of characteristic of time-like circular geodesic can be easily seen from Fig. 5. Therefore we conclude that among all the circular geodesics, the null-circular geodesics is characterized by the fastest way to circle of black hole.

## 8 Conclusions

In this paper we have clarify some aspect about principle Lyapunov exponent, KS entropy and unstable null-circular geodesics. We have presented that the principle Lyapunov exponent and KS entropy can be express in terms of the equation of ISCO (innermost stable circular orbit). Also we have highlighted that Lyapunov exponent can be utilize to determine the instability of equatorial circular geodesics for both time-like and null-circular geodesics. We have explored the relation for null-circular geodesics that the angular frequency($Q_o$) is equal to inverse of impact parameter ($D_o$), which is general characteristic of any stationary spacetime. Also we computed the equation of ISCO in which the smallest real root give the radius of ISCO for rotating black hole and for each of the special cases black holes like Kerr–Newman, Kerr black hole, Reissner-Nordström black hole and Schwarzschild black hole. We showed that $r_\sigma = 6M$ is the radius of ISCO for Schwarzschild black hole.

We have computed the equations of Lyapunov exponent and reciprocal of critical exponent for both the cases of time-like and null-circular geodesics for rotating black hole and for special cases. We constructed the equation of marginally bound circular orbit(MBCO) and showed that $r_\sigma = 4M$ is the radius of MBCO for Schwarzschild black hole . We have also determined the ratio of angular frequency of null-circular geodesics to time-circular geodesics. In this ratio we clarify that null circular geodesics have largest angular frequency than the time circular geodesics, that is, $Q_o > Q_\sigma$. This characteristic of null-circular geodesic can be check by the Fig. 4 above, which is graphed by taking $M = 1$ and $c = 2$. But in the ratio of time period of null-circular geodesics to time-circular geodesics. We can see that time-like circular geodesics have taken more time than null-circular geodesics, that is, $T_{timelike} > T_{photon}$. This scenario has been plotted in the Fig. 5, which is graphed by taking different values of radius $r$. In fact, we may conclude that any stable time-circular geodesics other than the ISCO traverses slowly than the null-circular geodesics.

**Acknowledgements** FR is grateful to the Inter-University Centre for Astronomy and Astrophysics (IUCAA), Pune, India for providing Associateship programme under which a part of this work was carried out. We are thankful to the referee for his valuable suggestions.

**Data Availability Statement** This manuscript has no associated data or the data will not be deposited. [Authors' comment: It's a theoretical work where the graphs were generated analytically.]